\documentclass{ws-ijbc}
\usepackage{ws-rotating} % used only when sideways tables/figures are used
\begin{document}
       
\catchline{}{}{}{}{} % Publisher's Area please ignore

\markboth{Katsanikas et al}{The structure and evolution of confined tori near a
  Hamiltonian Hopf Bifurcation}

\title{The structure and evolution of confined tori near a
  Hamiltonian Hopf Bifurcation}
\author{M. KATSANIKAS}
\address{Research Center for Astronomy, Academy of Athens\\
  Soranou Efessiou 4,  GR-11527 Athens, Greece}
\address{Section of Astrophysics, Astronomy and Mechanics, \\Department of
  Physics, University of Athens, Greece\\ mkatsan@academyofathens.gr}
\author{P.A. PATSIS and G. CONTOPOULOS}
\address{Research Center for Astronomy, Academy of Athens\\
  Soranou Efessiou 4,  GR-11527 Athens, Greece\\
  patsis@academyofathens.gr, gcontop@academyofathens.gr }

\maketitle

\begin{history}
\received{(to be inserted by publisher)}
\end{history}

\begin{abstract}
We study the orbital behavior at the neighborhood of complex unstable periodic 
orbits in a 3D autonomous Hamiltonian system of galactic type. At a  
transition of a family of periodic orbits from stability to complex 
instability (also known as Hamiltonian Hopf Bifurcation) the four eigenvalues 
of the stable periodic orbits  move out of the unit circle. Then the periodic orbits
become  complex unstable. In this paper  we  first integrate initial 
conditions  close  to the  ones of a complex  unstable periodic  orbit, which 
is close to the  transition point. Then, we plot the consequents  of the  
corresponding orbit in a 4D surface of section. To visualize this surface of 
section we use the method of color and rotation [Patsis and Zachilas 1994]. We 
find that the  consequents  are  contained in 2D ``confined tori''. Then, we 
investigate the structure of the  phase space  in the  neighborhood of complex 
unstable  periodic orbits, which are further   away from the transition 
point. In these  cases we observe clouds  of points in the 4D surfaces of 
section. The transition between the two types of orbital behavior is abrupt. 
\end{abstract}

\keywords{Chaos and Dynamical Systems, 4D surfaces of section, Hopf Bifurcation, Galactic Dynamics}

\twocolumn{
\section{Introduction}
The aim of this paper is to study the orbital behavior at the neighborhood of
a complex unstable periodic orbit in a 3D autonomous Hamiltonian system of
galactic type. Complex instability is a  type of instability of periodic
orbits that appears  in Hamiltonian  systems of three or more  degrees of 
freedom. 
\par In order to study the dynamical behavior at the neighborhood of a complex
unstable periodic orbit  we use the method of surfaces of section
[Poincar\'{e} 1892], which has  many  applications to  Dynamical Astronomy 
[for a review see e.g. Contopoulos 2002]. A basic problem
in  Hamiltonian systems  of  three degrees  of freedom is the visualization  
of  the 4D surfaces of section. Let us assume the phase space 
of an autonomous Hamiltonian system, that has 6 dimensions, e.g. in Cartesian 
coordinates, $(x,y,z,\dot x,\dot y,\dot z)$. For  a  given  Jacobi constant  
a trajectory  lies  on  a  5D  manifold. In   this manifold  the  surface  of  
section  is  4D. This does  not  allow  us  to   visualize   directly  the  
surface  of  section. 
\par Patsis and Zachilas [1994] proposed a method to visualize 4D spaces of 
section. It is based on rotation of the 3D projections of the figures in order 
to understand the geometry  of the  formed structures  and on color for
understanding the distribution of the consequents  in the 4th dimension.
We use for this application the ``Mathematica'' package [Wolfram 1999].
We work in Cartesian coordinates and we consider the $y=0$ with $\dot y>0$
cross section. A set of three coordinates (e.g. $(x,\dot x, \dot z)$) are used
for the 3D projection, while the fourth coordinate (in our example $z$) will 
determine the color of the consequents. There is a normalization of the color
values in the [min($z$), max($z$)] interval, which is mapped to
[0,1]. Following the intrinsic ``Mathematica'' subroutines our viewpoint is
given in spherical coordinates. The  distance $d$ of the 
center of the figure from the observer  is given by 
``Mathematica''  in a special scaled  coordinate system, in which the longest 
side of the  bounding box has length 1. For all figures  we use  $d=1$. The 
method associates  the smooth  distribution  or  the mixing of colors, with 
specific types of dynamical  behavior in the 4th  dimension [Patsis and
  Zachilas 1994][see also Katsanikas and Patsis 2011].

\par  The calculation  of the linear stability of a periodic orbit follows  
the method of Broucke [1969] and Hadjidemetriou [1975]. We  first consider  
small deviations from its initial conditions and then integrate the orbit 
again to the next upward intersection. In this way a  4D map (Poincar\'{e} 
map) is established and relates the initial with the final point. The relation 
of the  final deviations of this neighboring orbit  from the periodic one, 
with the initially introduced deviations, can be written in  vector form as 
$\xi$ = M $\xi_{0}$. Here $\xi$ is the final deviation, $\xi_{0}$ is the 
initial  deviation, and  M is a $4\times 4$ matrix, called the  monodromy 
matrix. It can be shown, that the characteristic equation  can be  written in 
the form  $\lambda^4 + a \lambda^3 + \beta \lambda^2 + a \lambda +1 = 0 $. Its 
solutions  $\lambda_{i},\; i=1,2,3,4$, due to the symplectic identity of the  
monodromy  matrix, that obey the  relations $\lambda_{1} \lambda_{3}=1$ and  
$\lambda_{2} \lambda_{4}=1$  can be written as:
\begin{equation} 
\lambda_i, \frac{1}{\lambda_i} = \frac {- b_i \pm \sqrt{b_i^2 - 4}}{2}, i=1,2
\end{equation}
where 

\begin{equation}
b_{1, 2} = \frac {a \pm \sqrt{\Delta}} {2}
\end{equation}

and

\begin{equation}
\Delta = a^2 - 4 (\beta - 2)
\end{equation}

\par The quantities $b_{1}$ and $b_{2}$ are called the stability indices. 
Following the notation of Contopoulos and Magnenat [1985], if  
$\Delta > 0,\; |b_1| < 2$  and  $|b_2| < 2$, all four eigenvalues are complex 
on the unit circle and the periodic orbit is called ``stable'' (S). If 
$\Delta > 0$  and $|b_1| > 2, \; |b_2| < 2$ or $|b_1| < 2,
\;|b_2| > 2$, the periodic orbit is called ``simple unstable" (U). In this case 
two eigenvalues are on the real axis and two are complex on the unit circle. 
If $\Delta > 0$  and $|b_1| > 2$  and  $|b_2| > 2$, the periodic orbit is 
called ``double unstable" (DU) and the four eigenvalues are on the real axis. 
Finally, if $\Delta < 0$   the periodic orbit is called ``complex unstable'' 
($\Delta$). In this case the  four eigenvalues are  complex numbers  and they 
are off the unit circle.  For the generalization of this kind of instability 
in Hamiltonian systems of N  degrees of freedom  the reader  may  refer to 
Skokos [2001]. If we have a stable one-parameter (in our system the 
Jacobi constant) family  of periodic orbits, the four eigenvalues of a stable 
periodic orbit are complex on the unit circle. By  varying  the parameter 
we have a pairwise collision of eigenvalues on two conjugate  points of the 
unit circle. From the Krein-Moser  theorem [e.g. Contopoulos 2002 p.298] we 
can decide if  after the collision of the  eigenvalues  they  will remain 
on the unit circle and the periodic orbits of the family  will stay stable, or  
if the eigenvalues  will  move out from the unit circle into the complex plane  
forming a  complex quadruplet. In this latter case the periodic orbits of the 
family will  become complex unstable and  we will have a  transition from 
stability to complex  instability (also known as Hamiltonian  Hopf
Bifurcation). From an analytical  point of view, the transition to  complex 
instability has been  studied using the Hamiltonian itself [Heggie 1985, Broer
  et al. 2007, Oll\'{e} et al. 2008] or 4D  symplectic maps [Bridges et al. 
1995]. In both cases, the approach  consists of  normal forms techniques 
[Oll\'{e} et al. 2005a,b] to simplify the Hamiltonian (or the map) and 
describe the local phase space structure near the  critical periodic orbit
(or fixed point in the  discrete context). Such analysis shows, that the 
transition to complex instability gives rise to bifurcating invariant 2D tori 
in the flow context or invariant curves in 4D  symplectic maps 
(as Poincar\'{e} map). The numerical computation of these 
\begin{figure}
\begin{center}
\includegraphics[scale=0.5]{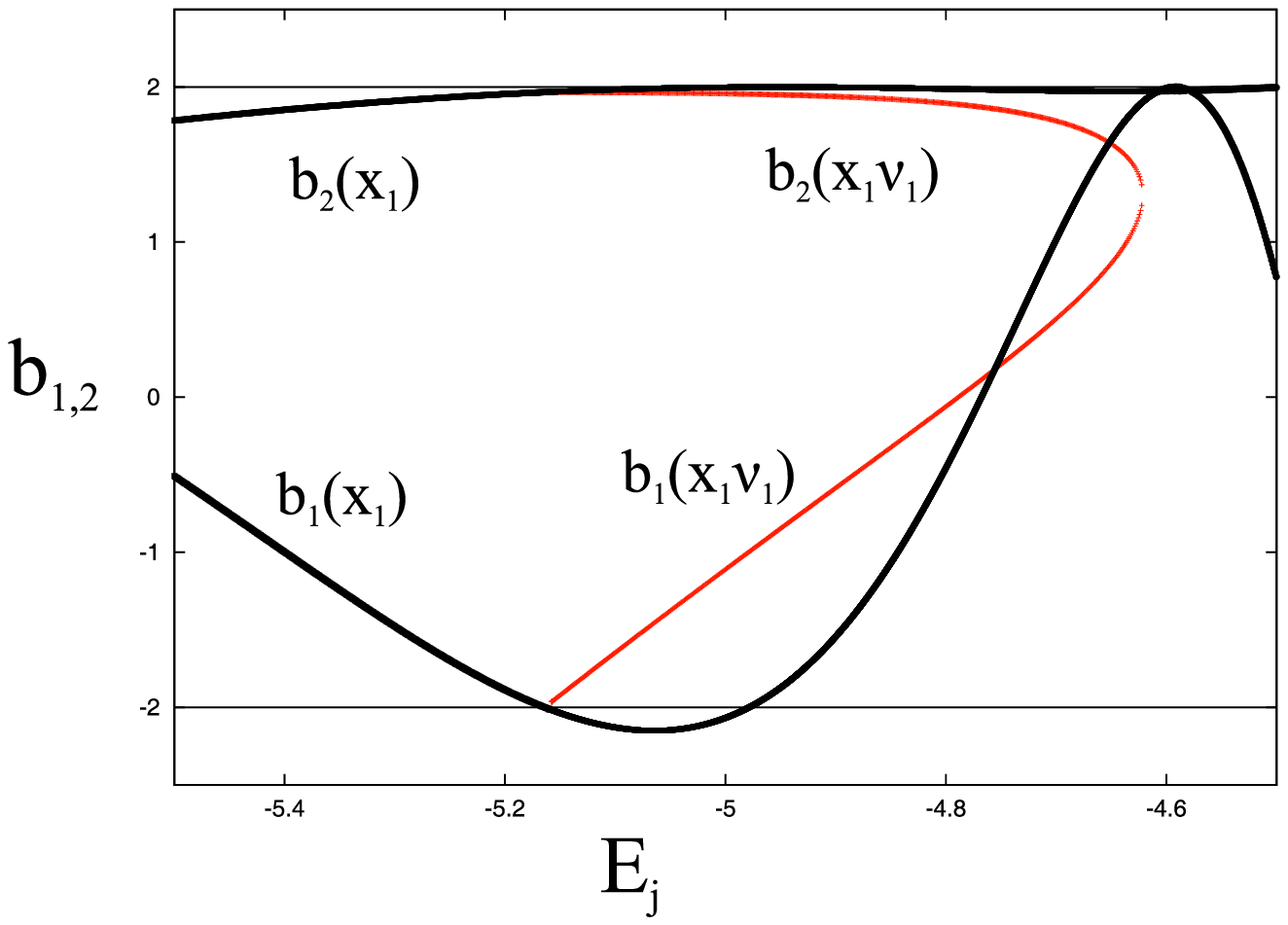}
\caption{Stability diagram  for $-5.5 < E_j < -4.5$, that shows the stability of
  family x1 (with black line) and its bifurcating family of p.o. x1v1 (with red
  line).}
\label{cstab1}
\end{center}
\end{figure}
invariant objects has been done  for Hamiltonian systems of three degrees of 
freedom [Pfenniger 1985b, Oll\'{e} and Pfenniger 1998, Oll\'{e} et al. 2004] 
and for 4D symplectic maps [Pfenniger 1985a, Jorba and Oll\'{e} 2004]. It is 
remarkable, that there exists not one but two kinds of Hamiltonian Hopf 
bifurcations (as it happens in the usual dissipative setting), depending on 
the coefficients of the normal form [Van der Meer 1985]. These two kinds of 
bifurcations are  usually called direct (supercritical) and 
inverse (subcritical). From a  numerical point of view we can distinguish the 
two kinds of Hamiltonian Hopf  bifurcation if we take  firstly  initial 
conditions in the vicinity of a  complex  unstable periodic orbit near the 
transition point from stability to complex instability. After 
that, we plot the consequents of the corresponding orbit  in the surface of 
section. If the consequents are  confined, we have a direct Hamiltonian Hopf 
Bifurcation. Otherwise, if the orbit escapes we have an inverse Hamiltonian 
Hopf Bifurcation. In this paper we examine the structure of the invariant 
surfaces  in the 4D surface of  section  in the neighborhood of a  complex 
unstable  periodic orbit after a  direct Hamiltonian Hopf bifurcation. The inverse Hopf bifurcation is not considered in the present paper, because in such a case all orbits close to a complex unstable periodic orbit escape [Jorba and Oll\'{e} 2004] and no confined structures appear.

\section{The  Hamiltonian System}
\label{sec:1}
The  system   we  use  for  our  applications  is described in details in Katsanikas and Patsis [2011]. It rotates  around its z-axis  with  angular  velocity  $\Omega_b= 60 \; km \; s^{-1} \; kpc^{-1}$. 
The Hamiltonian of our system is:

\begin{eqnarray}
H(x, y, z, \dot x, \dot y, \dot z)=
\nonumber\\ 
\frac {1}{2}(\dot x^2 + \dot y^2 + \dot z^2) + \Phi(x, y, z)
\nonumber\\
- \frac{1}{2} \Omega_b^2 (x^2 + y^2),
\end{eqnarray}
where  $\Phi(x,y,z)$ is our potential.
The  potential  $\Phi(x,y,z)$   in  its  axisymmetric  form  
can  be  considered as  a  representation of the  Milky  
Way approximated  by  two Miyamoto disks with masses $M_1$ and $M_2$ 
respectively  [Miyamoto and Nagai 1975].  The parameters we use in the present 
study are the same as in Katsanikas and Patsis [2011]. In  our  units, the  
distance  $R$=1  corresponds  to  1 kpc. The velocity unit corresponds to 
209.64~$km/sec$. For the Jacobi  constant (hereafter called the ``energy'')  
$E_j$=1  corresponds  to 43950 $(km/sec)^2$, while the velocity unit is 
209.64 $km/sec$.

\section{Spaces of section }
\label{sec:2}
\par A method to follow the stability  of a family of periodic orbits  in a 
system is by means of the ``stability diagram" [Contopoulos and Barbanis 1985, 
Pfenniger 1985a]. The stability  diagram gives the evolution of the stability 
of a  family of periodic orbits  in a system as one parameter varies, by means 
of the  evolution of the  stability indices $b_1, b_2$. In our case the 
parameter that varies is the energy $E_j$. Fig.~\ref{cstab1} gives the 
evolution of the stability of the central family of periodic orbits x1 on the 
equatorial plane [Contopoulos and Papayannopoulos 1980], and its  bifurcations 
for $-5.5 <$ $E_j < -4.5$. We observe, that x1 is  initially  stable and  
at $E_j$= $-$5.1644 it becomes simple unstable. There we have a $S \rightarrow 
U$ transition and a  new 3D family, x1v1 [Skokos et al 2002a,b], is 
bifurcating and is stable. For $E_j$=$-4.62$ the family x1v1 becomes complex  
unstable and the  stability  indices meet each other. After that the values of 
$b_{1,2}$ are complex numbers.

\begin{figure}
\begin{center}
\includegraphics [scale=0.5]{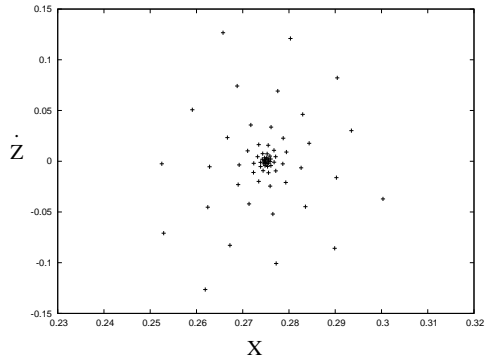}
\caption{2D projection $(x,\dot z)$ of the 4D  surface  of  section  at  the 
  neighborhood of   a  complex  unstable {\bf $(\Delta)$} periodic orbit  
  of the  3D  family  x1v1  for  $E_j=-4.619$ (for 70 intersections).}
\label{cproj}
\end{center}
\end{figure}

\par We investigate the structure of the phase space by perturbing the initial
conditions of a complex unstable periodic orbit in the 4D surface of section 
after a transition from stability to complex instability. For example we can 
apply  a perturbation $\Delta x = 10^{-4}$ at the initial conditions of a 
complex unstable periodic  orbit for $E_j$=$-4.619$  with initial conditions 
$(x_0, \dot x_0, z_0, \dot z_0)=(0.275137727,0,0.359838,0)$. Then, the $x$
initial condition  will be $x_0+\Delta x$.   
From  previous  papers [Contopoulos et al. 1994, Papadaki et al. 1995] we 
know that, for a number of intersections, a  spiral appears  in  the 
consequents  at  the  neighborhood  of  a  complex  unstable {\bf $(\Delta)$}  
periodic  orbit in 2D projections of the 4D  surfaces of  section. This 
structure  is  encountered  also  in  the  case  we  study
(Fig.~\ref{cproj}). We observe that the consequents are confined and  this 
means that we have a direct Hamiltonian Hopf bifurcation.
 \begin{figure}[t]
\begin{center}
\includegraphics[scale=0.5]{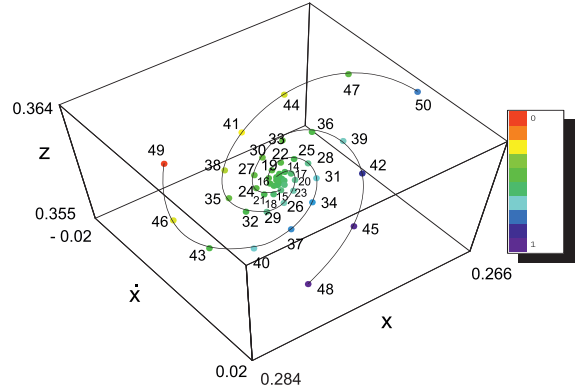}
\caption{The  orbital  behavior close  to  a  complex unstable  
  {\bf$(\Delta)$} periodic  orbit  of the  3D  family  x1v1 for 
  $\Delta x =10^{-4}$  and $E_j =-4.619$ (for 50  intersections). We  use 
  the  $(x,\dot x,z)$ space  for  plotting  the  points and  the  $\dot z$  
  value  to  color  them. Our  point  of  view  in  spherical coordinates is  
  given  by $(\theta, \phi) = (32^{o}, 53^{o})$.}
\label{csur1}
\end{center}
\end{figure}

\begin{figure}
\begin{center}
\includegraphics[scale=0.45]{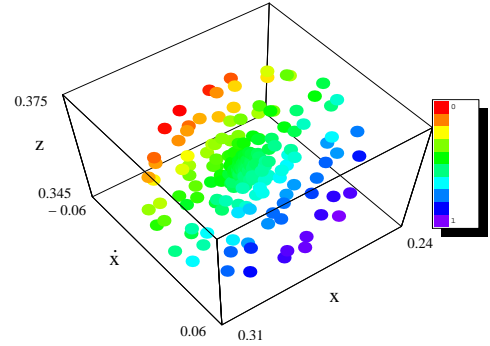}
\caption{The same  case  as  Fig. \ref{csur1}  for 2000  intersections.}
\label{csur2}
\end{center}
\end{figure}

\par We apply the method of color and rotation to study the invariant surface
in the neighborhood of the complex unstable periodic orbit. In Fig.~\ref{csur1}
we present the orbital behavior close to the complex unstable periodic orbit 
for 50 intersections with the surface of section. We use the 3D $(x,\dot x,z)$
projection for plotting the points and the $\dot z$ values to color them. Our 
point of view is  $(\theta, \phi) = (32^{o}, 53^{o})$. We observe a spiral
(with three spiral arms) in the neighborhood of the complex unstable periodic 
orbit. The first 13 points are  very close to the center of the spiral and
are not discernible in the figure. The 14th  point  is  located  on  a  
spiral  arm, the  15th  point is  located  on another  spiral arm and the 
16th  point  is  located on  the  third  spiral  arm. The 17th point is on the 
first spiral arm again  and so on. In this  way  the  points  fill the  space. 
Along every spiral arm we have a succession of colors (Fig.~\ref{csur1}). In 
Fig.~\ref{csur1} we observe that  if we begin from the center along the spiral 
arm  going through the 15th point we have a succession of colors from green to 
light blue, light blue to blue and blue to violet. Along the spiral arm 
through the 16th point we observe a smooth color variation from green to
light blue, to blue,  to light blue, to green, 
to green-yellow, to orange-red. Finally along the third 
spiral arm we observe also a  smooth color variation. Starting from the center 
of this spiral arm,  we see that green becomes green-yellow, then yellow, 
green-yellow, green and  finally light blue (Fig.~\ref{csur1}).

\par Figure \ref{csur2} depicts  the same case as in Fig.~\ref{csur1} and 
we observe the orbital behavior close to the complex  unstable periodic orbit 
for 2000 intersections with the surface of section. We cannot distinguish the
spiral any  more, but we  observe  a  disk-like structure. The central area is saturated due to the presence of many points. This object has
a very small thickness and is also slightly warped. Hereafter we will refer to
this structure as the ``disk'' or ``disk structure''. In the terminology of
Jorba \& Oll\'{e} [2004] such structures are called confined tori. Small 
deviations from  the pure planar geometry are due to a warp of the disk in the 
3D space. On this  disk  structure, from left to right, we have  a  
succession  of colors  from  red to orange, orange to  yellow, yellow to 
green, green  to light blue, light blue to blue and blue to  violet. For 10000 
intersections we observe that the  points fill the disk structure 
(Fig.~\ref{csur3}), which has a smooth color variation. The smallest  
distance of the  consequents on this disk structure  from the  complex
unstable periodic orbit is 0.000042 and the largest  distance is 0.059764 in  
the  3D projection $(x,\dot x ,z)$ of the 4D surface of section. The 
consequents  reach  an outermost  distance, then they move inwards reaching  a 
minimum  distance, then they move outwards etc. Some values for the minimum
and maximum distances are given  in Tables 1 and 2. We underline the fact, 
that the disk has an internal three-armed spiral structure. This means that if 
we choose arbitrarily a point on the disk, the subsequent consequents will 
follow a spiral pattern as the one presented in  Figs.~\ref{csur1},~\ref{csur2} and \ref{csur3}, which lies on the disk.
% In that sense the manifolds will coincide with the disk structure in this case. 
For this orbit we calculated also the ``finite time'' Lyapunov Characteristic Number ($LCN$), i.e.
\begin{displaymath}
LCN(t)=\frac{1}{t}\ln\left|\frac{\xi(t)}{\xi(t_0)}\right|,
\end{displaymath}
where $\xi(t_0)$ and $\xi(t)$ are the distances between two points of two 
nearby orbits at times t = 0 and t respectively( see e.g. Skokos [2010]). We found, that initially it decreases and finally it levels off around $6 \times 10^{-3}$ after about 4500 intersections (Fig.~\ref{peritto}, lower blue curve).

\begin{figure}
\begin{center}
\includegraphics[scale=0.4]{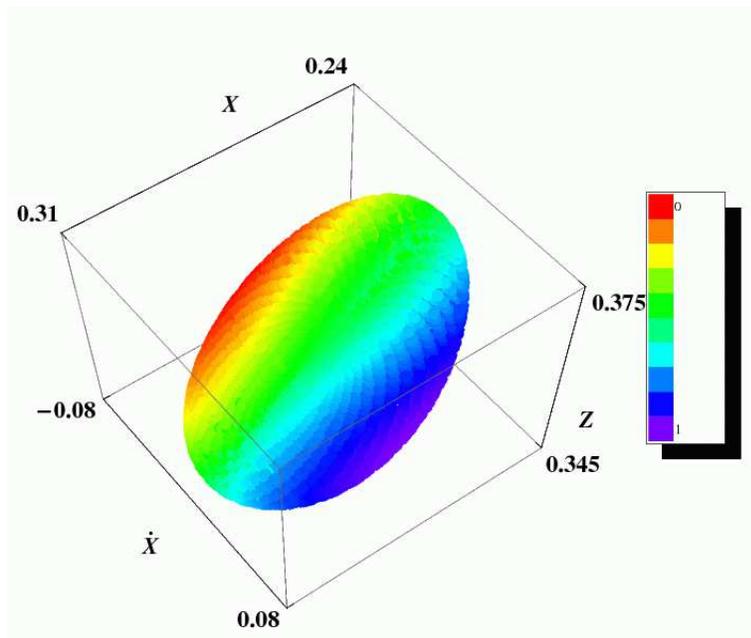}
\caption{The same  case  as  Fig. \ref{csur1}  for 10000  intersections.}
\label{csur3}
\end{center}
\end{figure}

\begin{figure}
\begin{center}
\includegraphics[width=8.65cm]{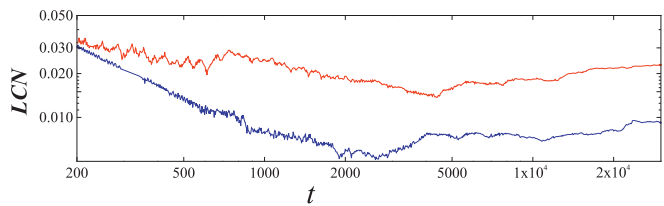}
\caption{The evolution of the two $LCN(t)$. The blue curve corresponds to the
  case of the confined torus, while the red to the case of the cloud. The axes
are in logarithmic scale.}
\label{peritto}
\end{center}
\end{figure}
  
\begin{figure}
\begin{center}
\includegraphics[scale=0.35]{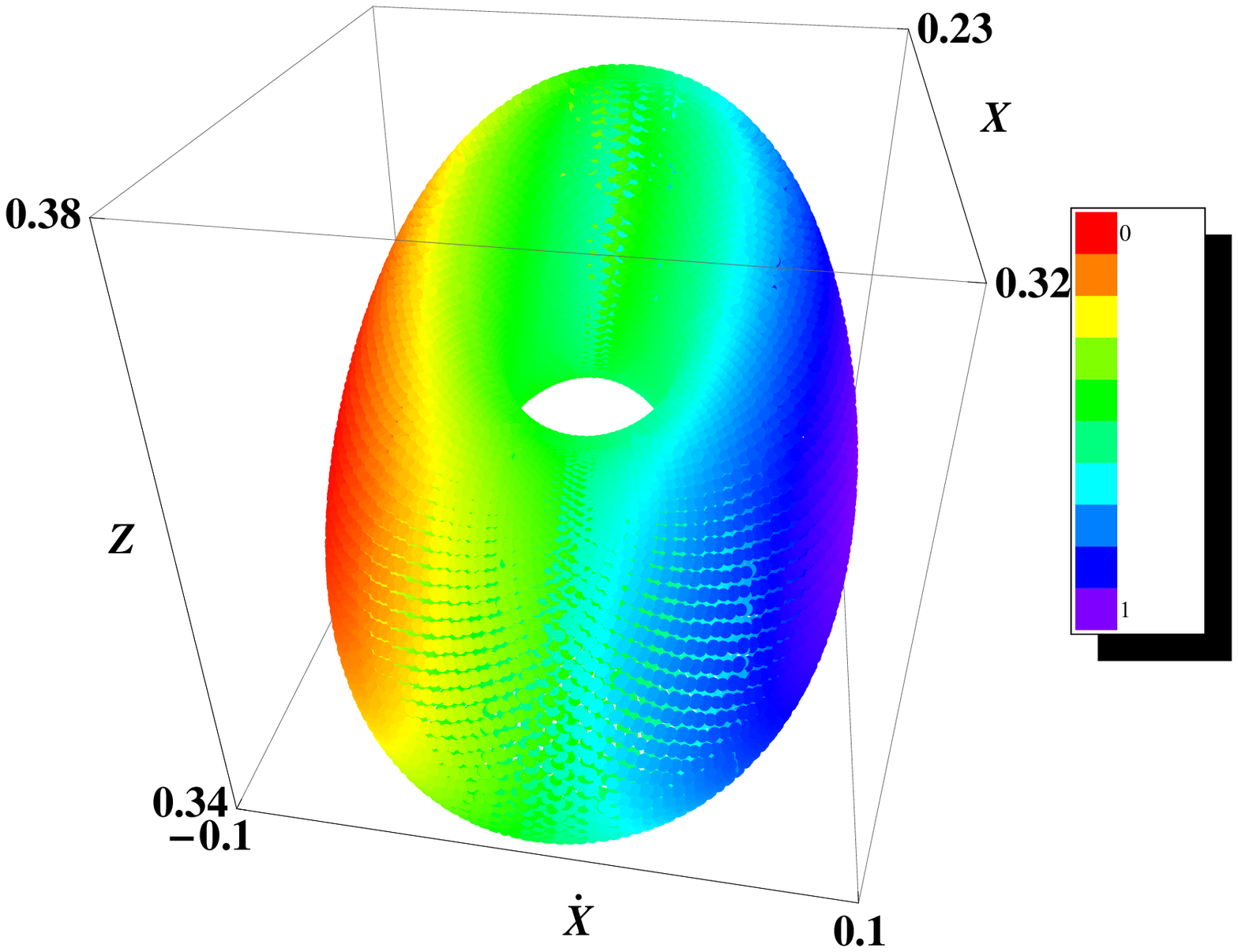}
\caption{The  orbital  behavior close  to  a  complex unstable  
  {\bf$(\Delta)$} periodic  orbit  of the  3D  family  x1v1 
  for $\Delta x =10^{-2}$  and  $E_j =-4.619$ (for 10000 intersections).  Our  
  point  of  view  in  spherical coordinates is  given  by   
  $(\theta, \phi) = (15^{o}, 90^{o})$.}
\label{disc1}
\end{center}
\end{figure}

\begin{table*}
\begin{center}
\tbl{Some examples of points on the  disk for a perturbation 
$\Delta x = 10^{-4}$ from the complex unstable periodic orbit for $E_j= -4.619$,
which have maximum  distance from the periodic orbit.}
{\begin{tabular}{|c|c|c|c|c|c|}\hline
nth point & 221& 522 & 802 & 2679& 8904\\ \cline{1-6}
max distance & 0.056284 & 0.058844 & 0.058453  & 0.059693 & 0.059764\\ \hline
\end{tabular}}
\end{center}
\end{table*}

\begin{table*}
\begin{center}
\tbl{Some examples of points on the  disk for a perturbation 
$\Delta x = 10^{-4}$ from the complex unstable periodic orbit for $E_j= -4.619$,
which have  minimum distance from the periodic orbit.}
{\begin{tabular}{|c|c|c|c|c|c|} \hline 
nth point& 300&600&912&2780&8990\\ \cline{1-6} 
min distance&0.000338& 0.000316&0.000042&0.000050& 0.000173\\ \hline 
\end{tabular}}
\end{center}
\end{table*}

\begin{figure}
\begin{center}
\includegraphics[scale=0.4]{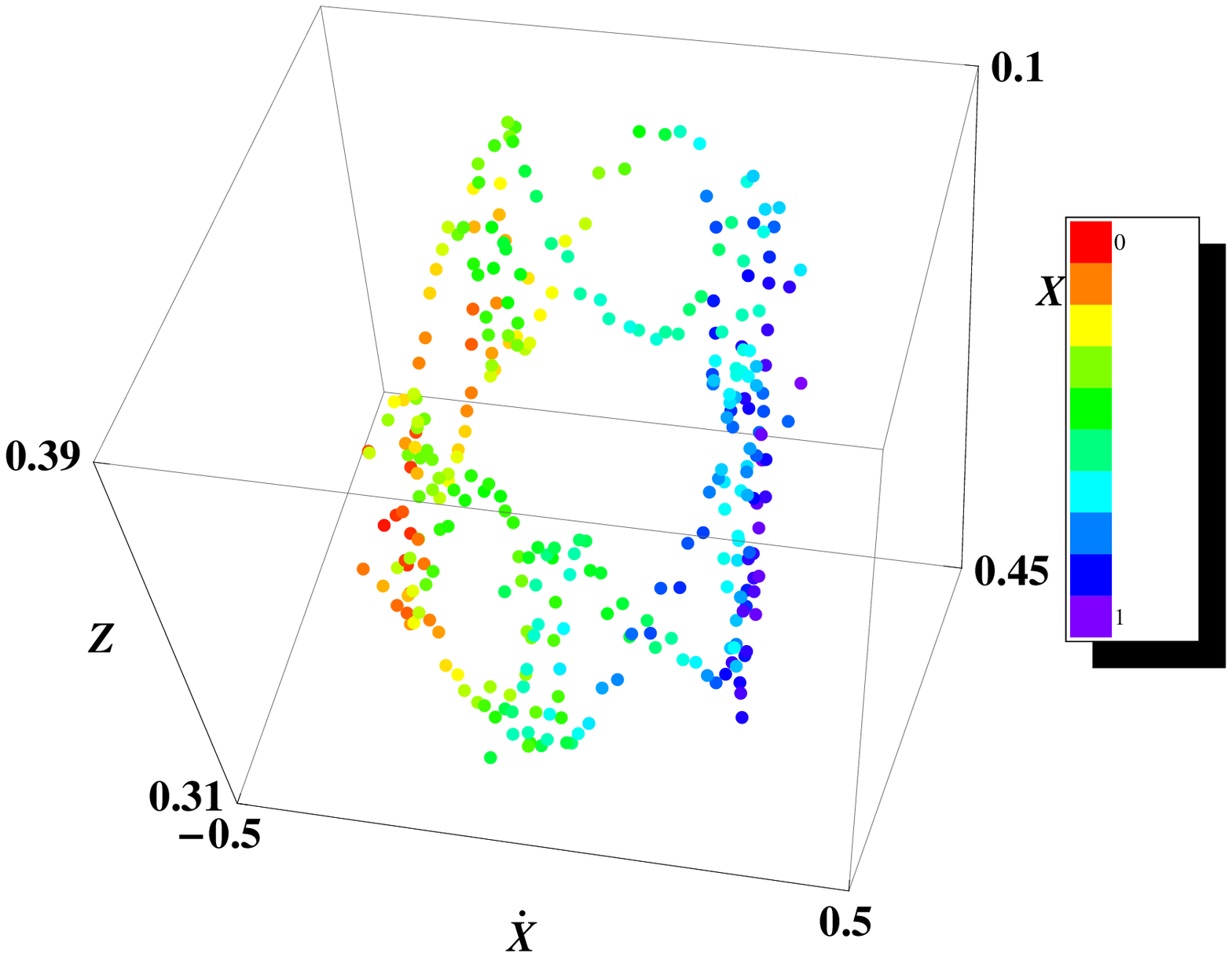}
\hspace{-4mm}
\caption{The  orbital  behavior close  to  a  complex unstable  
  {\bf$(\Delta)$} periodic  orbit  of the  3D  family  x1v1  for 
  $\Delta x =10^{-1}$  and  $E_j =-4.619$ (for 310 
  intersections).  Our  point  of  view  in  spherical coordinates is  given  
  by   $(\theta, \phi) = (15^{o}, 90^{o})$.}
\label{stick1}
\end{center}
\end{figure}

\begin{figure}
\begin{center}
\includegraphics[scale=0.65]{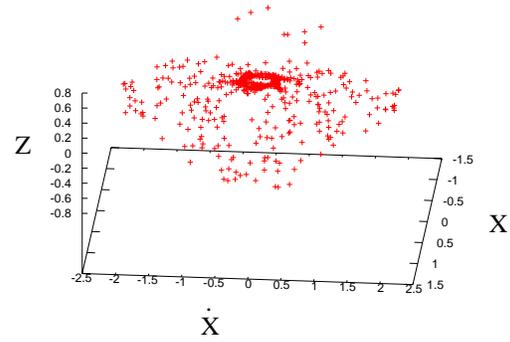}
\caption{The 3D projection  $(x,\dot x,z)$  of the consequents close  to  a  
  complex unstable {\bf$(\Delta)$} periodic  orbit  of the  3D  family  x1v1 
  for $\Delta x =10^{-1}$  and  $E_j =-4.619$ (for 3000 intersections).  Our  
  point  of  view  in  spherical coordinates is  given  by   
  $(\theta, \phi) = (55^{o}, 95^{o})$.}
\label{stick2}
\end{center}
\end{figure}

\par  We observe the same behavior (the same disk structure) if we  apply  a 
perturbation $\Delta x = 10^{-3}$ to the initial conditions of the  complex 
unstable periodic  orbit for $E_j$=$-4.619$ in the same way we did in the 
previous case for $\Delta x = 10^{-4}$. The only difference we find is that the 
minimum distance of the consequents from the  complex unstable periodic orbit 
is now 0.000999 and the maximum distance is 0.0592138. The minimum distance  is 
24 times larger than  the minimum distance for a perturbation 
$\Delta x =10^{-4}$ and for this reason we have a small hole at the center of 
the disk structure. Now if the perturbation is even larger, $\Delta x = 
10^{-2}$,  we observe  a different  disk structure (Fig. \ref{disc1}). The new 
disk becomes thicker than the previous disk we have studied with initial 
$\Delta x=10^{-3}$ or $\Delta x=10^{-4}$ away from the periodic orbit.  Now  
the ``thickness'' in the three coordinates is $(\Delta X , \Delta \dot X, 
\Delta Z)$ $=(0.09,0.2,0.04)$ instead of  $(\Delta X, \Delta \dot X, 
\Delta Z)$ $=(0.07,0.16,0.03)$ we had before. On  this ``disk structure''  
we  observe a  smooth color variation from  red to violet (Fig. \ref{disc1}). 
This means  that the 4th dimension follows the  topology  of  this  structure 
in the  4D surface of section. The smallest distance of the consequents on 
this  structure from the  complex unstable periodic orbit is 0.009999  and the 
largest  distance is 0.078655 in  the 3D  projection  $(x,\dot x ,z)$ of the 
4D surface of section. Thus, the smallest  distance of the consequents  on the 
disk structure for the  perturbation $\Delta x =10^{-2}$ is now 238 times 
larger than  the  smallest distance of the $\Delta x=10^{-4}$ perturbation. As 
expected the hole in middle of the ``disk'' is now even larger 
(Fig.~\ref{disc1}).

\par Now we increase the perturbation by taking  $\Delta x = 10^{-1}$, always 
for $E_j$=$-4.619$. The first approximately N=300 consequents form a toroidal surface. During this period the ``finite time'' $LCN(t)$ of the orbit decreases to a value $1.75 \times 10^{-2}$ (Fig.~\ref{peritto}, upper, red curve). Beyond that point it fluctuates as the time increases and finally increases and tends to level off around $2.5 \times 10^{-2}$ after about 4700 intersections. On our 4D surface of section the consequents fill a cloud around the toroidal object during the same time.

\begin{figure}
\begin{center}
\includegraphics[scale=0.85]{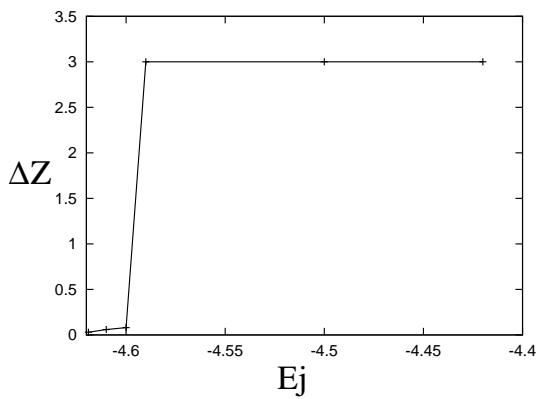}
\caption{Diagram  of the $\Delta$Z variation  of the  consequents close to a 
  complex unstable periodic orbit versus $E_j$.}
\label{dia}
\end{center}
\end{figure}

In Fig.~\ref{stick1} we observe the first 310 consequents. Besides the points that form the toroidal object, we have included also the very first points that depart from it. These points correspond to the cuspy features we can observe in the red dense ring in Fig.~\ref{stick2}. In Fig.~\ref{stick1} the 4th dimension $\dot z$ is represented by the 
colors. Smooth color variation in the consequents is again present and 
this means that we have a toroidal surface in the 4D space of section. By
considering  more points on the surface of section (Fig.~\ref{stick2}), we observe that they deviate 
from this toroidal surface and they soon occupy a large volume of the phase 
space. This indicates stickiness [Contopoulos and Harsoula 2008].

As mentioned previously, the maximal $LCN$ for the case of the cloud (Fig.~\ref{stick2}) is  around
$2.5 \times 10^{-2}$, while the same index for the orbit of the confined torus
(Fig.~\ref{csur3}) is lower, $6 \times 10^{-3}$. This difference is expected by the nature 
of the two orbits. In the cloud we have scattered points in 4 dimensions, while in the case of
the confined torus, chaos is due to the different maximum and
minimum distances from the center reached as the orbit ``fills" the disky
structure of this torus (see Tables 1 and 2). 
%Figure~\ref{peritto} shows the evolution of the two $LCN(t)$. With blue we depict the evolution of this index for the case of %the confined torus, while with red the corresponding index for the cloud.

\begin{figure}
\begin{center}
\includegraphics[scale=0.65]{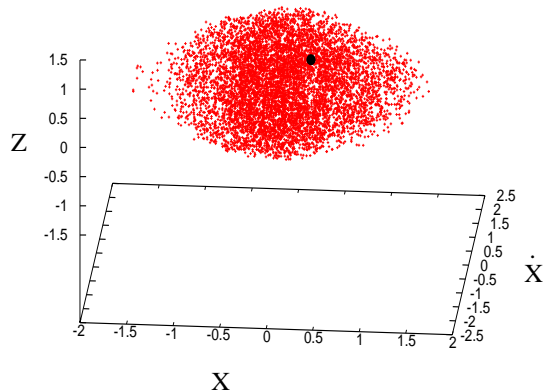}
\caption{The 3D projection  $(x,\dot x,z)$  of  the  consequents  close  to  a  
  complex unstable {\bf$(\Delta)$} periodic  orbit  of the  3D  family  x1v1 
   for $\Delta x =10^{-4}$  and  $E_j =-4.50$ (for 4000  intersections). The
   periodic orbit is depicted by a black point.  Our  point  of  view  in  
   spherical coordinates is  given  by  $(\theta, \phi) = (62^{o}, 5^{o})$.}
\label{cloud1}
\end{center}
\end{figure}

\begin{figure}
\begin{center}
\includegraphics[scale=0.4]{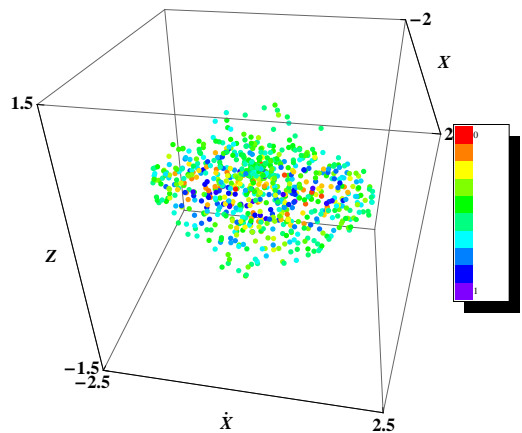}
\caption{The  orbital  behavior close  to  a  complex unstable  
  {\bf$(\Delta)$} periodic  orbit  of the  3D  family  x1v1 
  for $\Delta x =10^{-4}$  and  $E_j =-4.50$ (for 2000
  intersections).  Our  point  of  view  in  spherical coordinates is  given  
  by   $(\theta, \phi) = (55^{o}, 88^{o})$.}
\label{cloud2}
\end{center}
\end{figure}

\par  We choose now a $\Delta x=10^{-4}$ deviation from the initial conditions 
of the periodic orbit, that gives a confined torus at $E_j=-4.619$ and we 
increase  $E_j$. In Fig.~\ref{dia} we  depict the  variation of the thickness
$\Delta$Z of the  consequents in  the 4D surface of  section  versus the  
energy $E_j$. We  observe that for values of  $E_j$ between $-4.62$ and 
$-4.59$ we  have small values of $\Delta$Z  and  for  values of  $E_j$ between 
$-4.59$ and $-4.42$ we  have large values of $\Delta$Z. The first interval  
of $E_j$  corresponds to the disk structures that we described before. The 
second  interval corresponds to the clouds of points  that we find  for values 
of the  energy larger than $-4.59$. For example  for $E_j=-4.50$  we  observe 
a cloud  of points with  values of $z$  from $-1.5$ to 1.5 
(Fig.~\ref{cloud1}). The  periodic orbit is depicted as a black point and it 
has initial conditions  $(x_0, \dot x_0, z_0, \dot z_0)=$ 
$(0.307260,0,0.432148,0)$. In Fig.~\ref{cloud2} we observe a mixing of colors  
for this cloud and this indicates a chaotic behavior in 4D space.

\begin{figure}
\begin{center}
\includegraphics[scale=0.45]{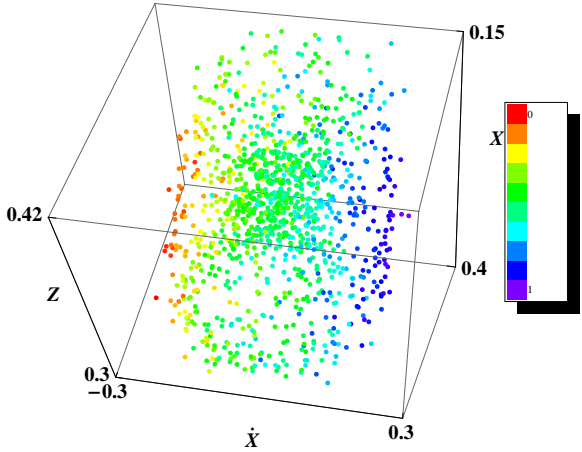}
\caption{The  orbital  behavior close  to  a  complex unstable  
  {\bf$(\Delta)$} periodic  orbit  of the  3D  family  x1v1 
  for $\Delta x =10^{-4}$  and  $E_j =-4.59$ (for 1250 intersections).  Our  
  point  of  view  in  spherical coordinates is  given  by   
  $(\theta, \phi) = (15^{o}, 90^{o})$.}
\label{stick3}
\end{center}
\end{figure}

\par Let us investigate now the changes observed in the neighborhood of the
complex unstable periodic orbit when we vary the energy.
At the transition point of Fig.~\ref{dia}, for $E_j=-4.59$  and for a 
perturbation of the initial conditions  equal to  $\Delta x = 10^{-4}$, the 
first 1250  consequents  remain on a  disk structure as we can see in
Fig.~\ref{stick3}. We note that the points do not suffice to fill densely the 
  ``disk''. On this disk  structure we observe a color succession from  
violet to blue, to light blue, to  green, to yellow, to orange, to  red. We 
underline  that the  succession of colors  means that the 4th dimension 
$\dot z$ of the  consequents lies on this disk structure. In Fig.~\ref{stick4} 
we see  that if we integrate the orbit  for more time, the  consequents  start  
to leave this disk  structure (that is depicted with black color) and form a  
cloud of points around it. This is again a typical behavior of a sticky 
orbit.

\begin{figure}
\begin{center}
\includegraphics[scale=0.65]{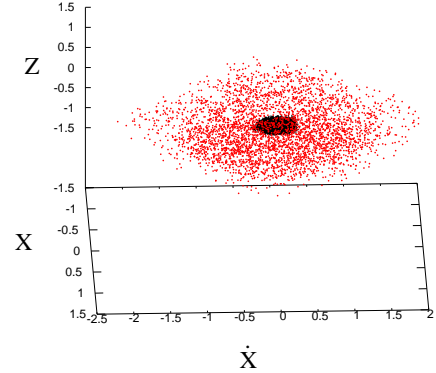}
\caption{The 3D projection  $(x,\dot x,z)$ of the consequents  close  to  a  
  complex unstable {\bf$(\Delta)$} periodic  orbit  of the  3D  family  x1v1 
  for $\Delta x =10^{-4}$  and  $E_j =-4.59$ (for 4000 intersections).  Our  
  point  of  view  in  spherical coordinates is  given  by   
  $(\theta, \phi) = (55^{o}, 88^{o})$.}
\label{stick4}
\end{center}
\end{figure}

\section{Conclusions}

In this paper we studied  the phase space structure  at the neighborhood of a
complex unstable periodic orbit after a direct Hamiltonian Hopf bifurcation. 
Close to a Hamiltonian Hopf bifurcation we observe that: 

\begin{enumerate}  

\item We have  a spiral structure formed by the consequents  at the 
  neighborhood of  complex unstable  periodic  orbits in the 2D  projections 
  of the 4D surface of section  as in Contopoulos et al [1994] and Papadaki et 
  al [1995]. Here we find that we have this spiral structure in the 3D 
  projections  and we observe a smooth color variation along their arms. 
  This means that  this spiral structure is a 4D object. 

\item The consequents near a complex unstable periodic orbit arrive  at a 
  maximum  distance from the  periodic orbit  and  they  move inwards. By
  repeating this process, a disk structure is formed in the  3D 
  projection of the  4D surface of section, which is called a confined torus
  [Pfenniger 1985a,b, Jorba \& Oll\'{e} 2004 and Oll\'{e} et al 2004]. On 
  this disk structure we  observe a smooth color succession. This  means that 
  we have a disk structure (the confined torus) also in the 4D space of 
  section.  

\item If we apply larger perturbations to the initial conditions 
  we observe, that the disk structures become toroidals with smooth  color  variation 
  and holes at their centers. For a critical value of the $\Delta x$
  perturbation of the initial conditions  we observe for a number of intersections a
  toroidal surface with smooth color variation. However, later,  the 
  consequents leave this toroidal surface  and move away occupying a larger 
  volume in the phase space. This is a case of stickiness 
  [Contopoulos \& Harsoula 2008] in the case  of complex instability and 
  confined tori. This is the first time that we  visualize stickiness 
  in the neighborhood of a complex unstable periodic orbit.

\item As the value of energy increases  we do not find confined tori anymore, 
  but clouds of  points in the neighborhood of complex unstable periodic
  orbits. These clouds have  mixing of colors. This means that we have strong 
  chaos in the 4D space of section when  we go  far from the  transition point 
  from stability to complex instability.

\item The calculation of the ``finite time'' $LCN$, gives a global value of the variation of the volume filled by the consequents of the orbits. We found that the $LCN(t)$ curve levels off after more than 4500 intersections and is around $2.5 \times 10^{-2}$ for the case of the cloud (Fig.~\ref{stick2}), and around $6 \times 10^{-3}$ for the confined torus (Fig.~\ref{csur3}). On the other hand, the method of color and rotation gives in a much more direct and detailed way the volumes of the phase space occupied by an orbit during its integration. 
This is of particular importance in Galactic Dynamics.
%, where we have phenomena like increase of the local dispersion of velocities %in a galaxy and we are interested about orbital behaviors in fractions of the Hubble time.
\end{enumerate}

\vspace{2cm}
\textit{Acknowledgments} MK is grateful to the ``Hellenic Center of Metals 
Research'' for its support in the frame of the current research. This research 
has been partly supported by the Research Committee of the Academy of Athens 
under the project 200/739.\\

\section{References}
  Bridges T.J., Cushman R.H. and Mackay R.S. [1995] ``Dynamics near an 
  irrational  collision of eigenvalues for symplectic mappings'' 
  \textit{Fields Inst. Comm.} \textbf{4}, 61-79.\\
  Broer H.W., Hanssmann H. and  Hoo J. [2007]  ``The quasi-periodic 
  Hamiltonian Hopf Bifurcation'' \textit{Nonlinearity} \textbf{20}, 417-460.\\
  Broucke R. [1969] ``Periodic orbits in the elliptic  restricted three-body 
  problem'' \textit{NASA Tech. Rep.} 32-1360, 1-125.\\
  Contopoulos G. and Papayannopoulos Th. [1980] ``Orbits in weak and strong 
  bars'' \textit{Astron. Astrophys.} \textbf{92}, 33-46.\\  
  Contopoulos G. and Barbanis B. [1985]  ``Resonant systems with three 
  degrees of freedom'' \textit{Astron. Astrophys.} \textbf{153}, 44-54.\\
  Contopoulos G. and Harsoula M. [2008] ``Stickiness in Chaos''
  \textit{Int. J. Bif. Chaos} \textbf{18}, 2929-2949.\\
  Contopoulos G. and Magnenat P. [1985] ``Simple three-dimensional  periodic 
  orbits in a  galactic-type potential'' \textit{Celest. Mech.} \textbf{37}, 
  387-414.\\ 
  Contopoulos G., Farantos S.C., Papadaki H. and  Polymilis C. [1994]  
  ``Complex unstable periodic orbits and their  manifestation in classical and 
  quantum dynamics'' \textit{Phys. Rev. E} \textbf{50}, 4399-4403.\\
  Contopoulos G. [2002] \textit{Order and Chaos in  Dynamical Astronomy}
  Springer-Verlag, New York Berlin Heidelberg.\\
  Hadjidemetriou J.D. [1975] ``The stability of periodic orbits in the 
  three-body problem''\textit{Celest. Mech.} \textbf{ 12}, 255-276.\\
  Heggie  D.C. [1985] ``Bifurcation at complex instability'' \textit{Celest. 
    Mech.} \textbf{35}, 357-382.\\  
  Jorba A. and Oll\'{e} M. [2004] ``Invariant curves near Hamiltonian Hopf 
  bifurcations of four-dimensional symplectic maps'' \textit{Nonlinearity} 
  \textbf{17}, 691-710.\\
  Katsanikas M. and Patsis P.A. [2011] ``The structure of invariant tori in a
  3D galactic potential'' \textit{Int. J. Bif. Chaos} (in press).\\
  Miyamoto M. and Nagai R. [1975] ``Three-dimensional models for the 
  distribution of mass in galaxies'' \textit{Publ. Astron. Soc. Japan} 
  \textbf{27}, 533-543.\\
  Oll\'{e} M. and  Pfenniger D. [1998] ``Vertical orbital structure around the 
  lagrangian points in barred galaxies. Link with the secular evolution of 
  galaxies'' \textit{Astron. Astrophys.} \textbf{334}, 829-839.\\
  Oll\'{e} M., Pacha J.R. and Villanueva J. [2004] ``Motion close to the Hopf 
  bifurcation of the vertical family of periodic orbits of $L_4$'' 
  \textit{Celest. Mech. Dyn. Astr.} \textbf{90}, 89-109.\\
  Oll\'{e} M., Pacha J.R. and Villanueva J. [2005a] ``Quantitative estimates 
  on the  normal form around a non-semi-simple 1:-1 resonant periodic orbit'' 
  \textit{Nonlinearity} \textbf{18}, 1141-1172.\\ 
  Oll\'{e} M., Pacha J.R. and Villanueva J. [2005b] ``Dynamics close to a 
  non-semi-simple 1:-1 resonant periodic  orbit'' \textit{Discrete Contin. 
    Dyn. Syst. B} \textbf{5}, 799-816.\\
  Oll\'{e} M., Pacha J.R. and Villanueva J. [2008] ``Kolmogorov-Arnold-Moser 
  aspects of the periodic Hamiltonian Hopf bifurcation'' \textit{Nonlinearity} 
  \textbf{21}, 1759-1811.\\
  Papadaki H., Contopoulos G. and Polymilis C. [1995] ``Complex Instability'' 
   In: \textit{From Newton to Chaos} ed by A.E. Roy, B.A. Steves, 
  Plenum Press, New York, pg. 485-494.\\
  Patsis P.A. and Zachilas L. [1994] ``Using Color and rotation for 
  visualizing  four-dimensional Poincar\'{e} cross-sections:with applications 
  to the orbital  behavior of a three-dimensional Hamiltonian system'' 
  \textit{Int. J. Bif. Chaos}  \textbf{4}, 1399-1424.\\
  Pfenniger D. [1985a] ``Numerical study of complex instability:I Mappings''  
  \textit{Astron. Astrophys.} \textbf{150}, 97-111. \\ 
  Pfenniger D. [1985b] ``Numerical study of complex instability:II 
  Barred galaxy bulges'' \textit{Astron. Astrophys.} \textbf{150}, 112-128.\\ 
  Poincar\'{e} H. [1892] \textit{Les M\'{e}thodes Nouvelles  de la  
    M\'{e}canique  C\'{e}leste} Gauthier Villars, Paris I (1892), II (1893), 
  III (1899); Dover (1957).\\
  Skokos Ch. [2001] ``On the stability of periodic orbits of high dimensional 
autonomous Hamiltonian systems'' \textit{Physica D} \textbf{159}, 155-179.\\
  Skokos Ch. [2010]   ``The Lyapunov Characteristic Exponents and their 
Computation'', \textit{Lect. Not. Phys.} \textbf{790}, 63-135.\\
  Skokos Ch., Patsis P.A. and  Athanassoula E. [2002a] ``Orbital dynamics of 
  three-dimensional bars-I. The backbone of three-dimensional bars. A fiducial 
  case'' \textit{Mon. Not. R. Astr. Soc.} \textbf{333}, 847-860.\\
  Skokos Ch., Patsis P.A. and Athanassoula E. [2002b] ``Orbital dynamics of 
  three-dimensional bars-II. Investigation of the parameter space''  
  \textit{Mon. Not. R. Astr. Soc.} \textbf {333}, 861-870.\\
  Van der Meer J-C. [1985] \textit{The Hamiltonian Hopf Bifurcation} 
  Lecture Notes in  Mathematics Vol. 1160, Berlin.\\
  Wolfram S. [1999] \textit{The Mathematica book} Wolfram media \&  Cambridge  
  Univ. Press. \\

\end{document}